\documentclass[conference]{IEEEtran}
\IEEEoverridecommandlockouts
\pdfoutput=1
\usepackage{cite}
\usepackage{amsmath,amssymb,amsfonts}
\usepackage{booktabs}
 \usepackage{enumitem}
\usepackage{threeparttable}
\usepackage{algorithmic}
\usepackage{graphicx}
\usepackage{textcomp}
\usepackage{xcolor}
\def\BibTeX{{\rm B\kern-.05em{\sc i\kern-.025em b}\kern-.08em
    T\kern-.1667em\lower.7ex\hbox{E}\kern-.125emX}}
\begin{document}

\title{Helper Recommendation with seniority control in Online Health Community}

\author{\IEEEauthorblockN{ Junruo Gao}
\IEEEauthorblockA{\textit{China Telecom Cloud Technology Co., Ltd}\\junruo.gao@gmail.com}
\and
\IEEEauthorblockN{Chen Ling}
\IEEEauthorblockA{\textit{Emory University} \\chen.ling@emory.edu
}
\and
\IEEEauthorblockN{Carl Yang}
\IEEEauthorblockA{\textit{Emory University} \\
j.carlyang@emory.edu}
\and
\IEEEauthorblockN{Liang Zhao}
\IEEEauthorblockA{\textit{Emory University} \\
liang.zhao@emory.edu}

}
\maketitle

%%
%% The abstract is a short summary of the work to be presented in the
%% article.
\begin{abstract}
 Online health communities (OHCs) are forums where patients with similar conditions communicate their experiences and provide moral support.
 Social support in OHCs plays a crucial role in easing and rehabilitating  patients. However, many time-sensitive questions from patients often remain unanswered due to the multitude of threads and the random nature of patient visits in OHCs.
 To address this issue, it is imperative to propose a recommender system that assists solution seekers in finding appropriate problem helpers.
Nevertheless, developing a recommendation algorithm to enhance social support in OHCs remains an under-explored area. Traditional recommender systems cannot be directly adapted due to the following obstacles.
First, unlike user-item links in traditional recommender systems, it is hard to model the social support behind helper-seeker links in OHCs since they are formed based on various heterogeneous reasons.
Second, it is difficult to distinguish the impact of historical activities in characterizing patients. 
Third, it is significantly challenging to ensure that the recommended helpers possess sufficient expertise to assist the seekers.
To tackle the aforementioned challenges, we  develop a  Monotonically
regularIzed diseNTangled Variational Autoencoders (MINT) model to strengthen social support in OHCs. Specifically, we formulate the interactions between seekers and helpers on specific threads as a dynamic graph, utilizing the encoded historical activities as node features. Additionally, we propose a graph-based disentangle VAE to capture both time-invariant and time-varying patient features.
To ensure the logical pairing of seekers and helpers, we introduce a monotonic regularizer to model patients' experiences and incorporate a seniority-level constraint to guarantee appropriate expertise in the recommended helpers.
Extensive experiments were conducted, and the results demonstrate the exceptional performance of our proposed method.
\end{abstract}

%%
%% The code below is generated by the tool at http://dl.acm.org/ccs.cfm.
%% Please copy and paste the code instead of the example below.
%%

%% A "teaser" image appears between the author and affiliation
%% information and the body of the document, and typically spans the
%% page.
% \begin{teaserfigure}
%   \includegraphics[width=\textwidth]{sampleteaser}
%   \caption{Seattle Mariners at Spring Training, 2010.}
%   \Description{Enjoying the baseball game from the third-base
%   seats. Ichiro Suzuki preparing to bat.}
%   \label{fig:teaser}
% \end{teaserfigure}

%%
%% This command processes the author and affiliation and title
%% information and builds the first part of the formatted document.
\maketitle

\section{Introduction}
 %Online health communities (OHCs) are online venues that primarily provide a means for patients and their families to learn about illnesses, seek and offer social support, and connect with others in similar circumstances []. 
%Distinct from general online communities, OHCs focus exclusively on health-related threads for those currently navigating the world of diseases, illness, and medicine [].  Furthermore, unlike health-related websites, etc. that only allow patients to retrieve information, OHCs that take many forms, such as blogs, chats, forums, and social media sites [] also allow for social communication  between multiple patients. 
%During various social support formed between seekers and helpers in OHCs, such as sharing illness experiences, and so on, there have been a large number of patients who obtain emotional support and increase disease-specific expertise [].
 
Overcoming a health issue is a challenging and emotional process. Obstacles often encompass an absence of understanding about the illness, difficulty in comprehending the severity of the disease as communicated by experts and feeling emotionally vulnerable due to loss of control \cite{johnston2013online}. Online health communities (OHCs) have emerged as a growing platform for patients and their families to acquire knowledge about illnesses, facilitate the exchange of information and experiences, and connect with individuals who have undergone similar health challenges \cite{mo2008exploring, meier2007cancer}. Unlike general online communities, OHCs can take diverse forms, including blogs, forums, and social media platforms \cite{van2013using}, enabling patients to engage in various forms of asynchronous social communication.

\begin{figure}
\centering
\includegraphics[width=\columnwidth]{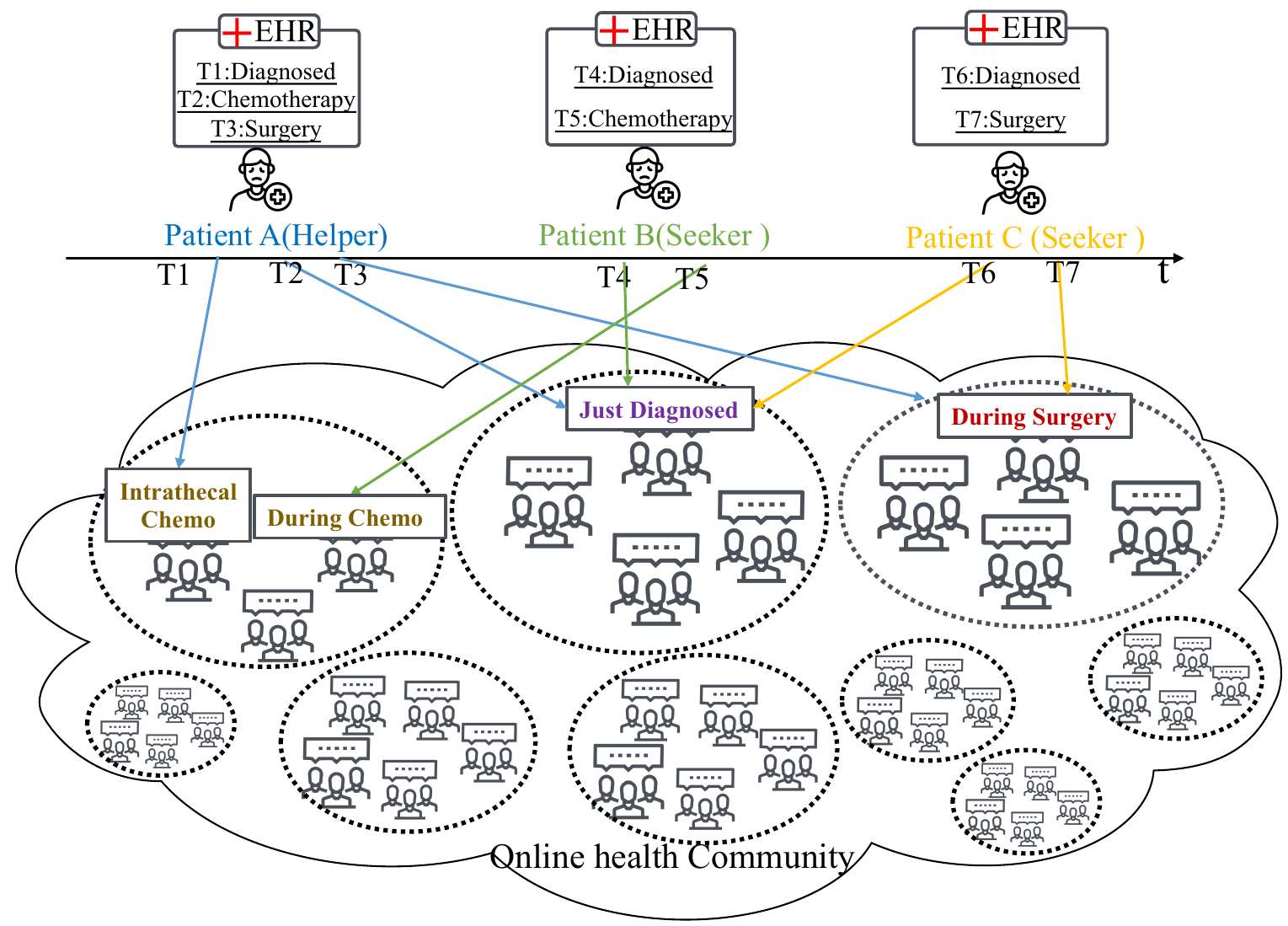}
\caption{The dynamic activities of patients in OHCs. Patient $A$ went through more health stages and visited more threads that would help Patient $B$ and Patient $C$ answer their questions on corresponding threads.}
\label{intro}
\vspace{-0.2cm} 
\end{figure}

Within OHCs, patients have access to various discussion threads where they can seek advice on illness-related questions or share their own experiences and emotions, e.g., \emph{During Surgery} and \emph{During Chemo} in Figure \ref{intro}. However, due to the asynchronous nature of OHCs and the randomness of patient visits, there is often a long waiting time for many questions to be answered. To enhance communication efficiency and effectiveness within OHCs, it is crucial to develop a recommender system that can connect patients seeking urgent health-related advice (i.e., solution seekers) with experienced patients or healthcare providers (i.e., problem helpers).

Most existing recommendation algorithms \cite{rendle2009bpr, lightgcn} have made success in modeling collaborative filtering relationships between users and items. However, recommending appropriate problem helpers to solution seekers requires capturing the affinity between users and their health stages, as well as their interactions with visited threads. These aspects go beyond what classical user-item algorithms can effectively capture. Another research avenue \cite{ning2019personet,cheng2019friend,li2020friend} is primarily focused on recommending friends within social networks. By leveraging existing social relationships, these approaches frame the recommendation problem as a link prediction task.  For instance, researchers \cite{sankar2021graph} propose to model the users' historical interests and build implicit relationships by computing the similarity for friendship recommendations. However, the helper recommendation task in OHCs cannot be adequately addressed as a simple link prediction problem, as a more intricate logical relationship exists between solution seekers and problem helpers. Specifically, recommended helpers in OHCs must possess specific experience with the relevant illness to effectively respond to the queries posed by seekers. Consequently, the helper recommendation task cannot be satisfactorily solved solely by the conventional link prediction problem. A more intricate logical relationship exists, where the suitability of a helper to address a seeker's questions depends on their particular experience.  For instance, as depicted in Figure \ref{intro}, even though both patient $B$ and patient $C$ have previously visited the thread \emph{Just Diagnosed} before and may appear similar, patient $B$ can not adequately address the questions posed by patient $C$ in the thread \emph{To do list for Surgery}.
%Various past work has already shown the effectiveness of the recommendation algorithms. Most of them mainly focus on modeling collaborative filtering relationships between   users and   items to provide recommendations, which  achieve great convenience and efficiency. However, in OHCs, patients' experienced health stages and historical activities  can greatly influence their interactions, which cannot be captured by classical algorithms. As Figure \ref{intro} shows, patient $A$, patient $B$ and patient $C$ all have gone through the health stage `Diagnosed', there is  a similarity between them, which leads to the later interactions. Another line of research work  mainly focuses on recommending  friends on social networks. Based on given social relationships, they consider the recommendation problem as a link prediction task. Researchers in [] propose to model historical users' interests and build an implicit relationship between users by computing the similarity between users to achieve recommendations. But they can ignore the logical relation between seekers and helpers in OHCs, namely, the recommended helpers need to be experienced enough to answer seekers' questions. For example, as shown in Figure \ref{intro}, although patient $B$ and patient $C$ both visited the thread `Just Diagnosed' before and  may be considered similar, patient $B$ can not answer the questions on the thread `To do list for Surgery' proposed by patient $C$.

Given the distinctive recommendation scenario in OHCs, the straightforward adaptation of existing recommendation algorithms for the purpose of pairing ``problem helpers'' with ``solution seekers'' faces several significant obstacles. \textbf{1). Difficulty in modeling the implicit interaction between problem helpers and solution seekers due to heterogeneous factors.}
Unlike traditional user-item recommender systems, the interactions between problem helpers and solution seekers in OHCs are influenced by diverse and heterogeneous factors, including historical activities and experienced health stages. These factors intricately shape the nature of patient interactions, making the modeling of social support within OHCs a complex task.
%Despite the effectiveness of the above solutions, none of the work is designed for seeker-helper pairing in OHCs, which cannot be effectively modeled by simply integrating the existing techniques, due to several significant challenges: 
%\textbf{1). Difficulty in  modeling the social support between problem helpers  and solution seekers.}
% In OHCs, seekers ask questions on specific threads, while helpers who have gone through similar health stages and have the related knowledge would answer   questions under the same threads. 
% Such the complexity of patient interactions based on their experienced health stages and visited threads makes it challenging to build a model to capture the social support between patients 
\textbf{2). Difficulty in distinguishing influence of historical activities in OHCs to comprehensively characterize patients.} Patients tend to have a combination of static and evolving features, encompassing inherent knowledge about their illness as well as dynamically acquired knowledge through their interactions with various threads over time. Properly discerning the impact of historical activities is crucial for comprehensive patient characterization, as it necessitates considering both time-invariant and time-varying features.  However, disentangling these features during the modeling process poses a significant challenge. \textbf{3). Difficulties in ensuring the competence of predicted helpers in addressing seekers' questions.}  In OHCs,  relying solely on similarity measures between patients is inadequate for accurate seeker-helper recommendations, as there exist logical seniority orders \cite{vydiswaran2019identifying} as Figure \ref{intro} shows. The existing recommendation algorithms are currently unable to incorporate such logical seniority  in a straightforward manner. Additionally, employing a simple filter based on primary patients is suboptimal for an end-to-end system, as different patients possess varying levels of expertise, making it time-consuming to filter for each seeker before making recommendations.

In order to address   the aforementioned challenges comprehensively, we develop a  novel approach called	\textbf{M}onotonically regular\textbf{I}zed dise\textbf{NT}angled    VAE (\textbf{MINT})  for the recommendation of problem helpers to solution seekers  in  OHCs. MINT is specifically designed to overcome the identified obstacles by incorporating three key components. Specifically, we formulate the interactions between seekers and helpers as a dynamic graph to deal with the first challenge and consider the modeled historical information as patient node features.   To tackle the second challenge,  we propose graph-based Disentangled VAE  to learn time-varying and time-invariant features. Lastly, to solve the third challenge, we design a monotonic regularization to enforce the correspondence between time-varying features and the senior level and leverage a senior constraint to guarantee the senior logic between seekers and predicted helpers.
\begin{itemize}[leftmargin=*]
\item \textbf{Building a novel framework to model the social support between patients based on a dynamic graph.}  This graph-based representation allows us to capture the complex and heterogeneous nature of social support in OHCs, with the modeled historical information serving as essential patient node features.
\item \textbf{Proposing a graph-based Disentangled VAE   to characterize the time-invariant and time-varying features.}  This disentanglement facilitates a comprehensive understanding of patients' characteristics, capturing the dynamic evolution of their knowledge while preserving the stable aspects of their expertise.
\item \textbf{Designing  a monotonic regularizer and a seniority level constraint to guarantee the logic between seekers and predicted helpers.} The monotonic regularizer enables the learned time-varying features monotonically correlate with the seniority level.  The seniority level constraint guarantees that the recommended helpers possess the necessary expertise to assist the seekers.
\item  \textbf{Conducting Extensive experiments to validate the effectiveness of  MINT.}   The results show the superiority of our proposed model over the comparison methods by 17\% and 9\% on the two real-world datasets. 
\end{itemize}

The rest of the paper is organized as follows. Section II
outlines related research work in Recommender Systems, online health communities analysis,  and VAE models. 
Section III describes in detail each component of the proposed MINT framework.
The problem formulation of the  dynamic social support graph, thread sequence, health stage sequence, and senior level sequence was introduced in Section III-A.
The objective function  of seeker-helper modeling was introduced in Section III-B, including  the evidence lower bound (ELBO) loss graph-based disentangled VAE, smoothness constraints, the Bayesian
Personalized Ranking loss,  monotonic regularization, and  a seniority-level constraint. In Section III-C, a graph-based disentangled variational autoencoder was proposed to model time-invariant and Time-varying features. In Section III-D, a novel recommending method was proposed under logical seniority orders for OHC patients.
Finally, we did an  amount of experimental analysis and
case studies in Section Section IV and demonstrated the effectiveness
of the proposed framework.

 \section{Related work}
\noindent\textbf{Recommender System.}
Traditionally, collaborative filtering techniques have been widely employed in recommender systems to generate recommendations based on user-item interactions. However, these approaches often suffer from data sparsity and cold-start problems, limiting their effectiveness. Graph Neural Networks (GNNs) offer a promising solution by effectively leveraging the underlying graph structure of the user-item interactions.
The key idea behind these approaches is to model the user-item interactions as a graph, where users and items are represented as nodes, and the interactions between them are represented as edges. By employing graph convolutional layers, GNNs can capture the high-order connectivity patterns present in the graph, allowing for better representation learning of users and items. 
Graph Neural Networks (GNNs) can be applied not only in static recommendation scenarios but also in sequential recommendation scenarios to capture the temporal dynamics of user preferences and item interactions.
% With the advancement of graph neural networks,  researchers in \cite{wang2019neural,lightgcn} exploit the graph structure to facilitate recommendations based on a user-item graph. 
By representing users and items as nodes in a graph and the sequential interactions as edges in the graph, GNN layers can learn the temporal dependencies and evolution of the nodes \cite{chang2021sequential,fan2021continuous,zhang2022dynamic,zhang2021tg}. This enables the model to effectively capture the evolution of user preferences and transitions between items.
Furthermore, in order to learn  different factors that influence user-item interactions,  disentangled graph methods are proposed. They   combine the principles of collaborative filtering, which utilizes user-item interactions, with disentanglement learning, which aims to disentangle underlying factors of variations in the data \cite{wang2020disentangled}. Zheng et.al proposes CLSR which further disentangles long and short-term interests for recommendation by separating and modeling the distinct preferences and behaviors of users that are either persistent over a long period or  temporary \cite{zheng2022disentangling}. This disentanglement enables more accurate and personalized recommendations that consider both the immediate interests and the long-term preferences of users. However, CLSR only considers the static profiles of users, ignoring the influence of dynamic  user features.
Instead of recommending specific products or items, friend recommendations, also known as social recommendations
 is another important aspect in the realm of online social media, as highlighted by  \cite{ning2019personet,cheng2019friend,li2020friend}.
The intricate information embedded in social networks can be harnessed to enhance the learning process by considering the social effects among users. This type of recommendation can enhance social interactions, foster community engagement, and expand users' social circles.
Although the method proposed in \cite{sankar2021graph} using graph-based method models user interests or features to compute user similarity  for matching purposes, they are not applicable to seeker-helper recommendation scenarios as they neglect to incorporate logical seniority orders among patients.

\textbf{Online Health Communities}:
Online health communities (OHCs) can naturally be formed as a heterogeneous network of user, patient, and discussion threads \cite{ling2021deep,ling2023motif}, which have offered platforms for acquiring health-related information.  There are various OHCs related to different diseases   \cite{forum, forum1, forum2}, and a substantial number of patients seeking healthcare assistance. To enhance the service provided to patients, several approaches have been developed to mine and analyze patients' activities within OHCs.  In \cite{zhao2014finding,chen2020exploring}, focus on capturing patients' sentiments expressed in their discussions to identify reliable medical knowledge. In order to further aid healthcare organizations in supporting patients,  Gao et al. \cite{gao2019dyngraph2seq,gao2022modeling} propose to infer patients’ health stages based on their online activities. In \cite{vydiswaran2019identifying}, helpers in OHCs tend to possess more specialized expertise compared to seekers and propose explicitly assessing their level of expertise. Despite its significance, few methods have addressed the problem of recommendation within OHCs.
The only work \cite{yang2018enriching}  relevant to OHC recommending proposes to recommend threads for patients. This work overlooks the crucial aspect of social support among patients, which plays a vital role in facilitating efficient helping and seeking.

\textbf{Variational Autoencoders}:
Variational Autoencoders (VAEs) are a popular class of generative models that have gained significant attention in the field of deep learning and machine learning. They combine the power of deep neural networks with probabilistic modeling to learn rich latent representations of input data and generate new samples. $\beta$-VAE introduces a hyperparameter $\beta$ to control the trade-off between reconstruction quality and disentanglement in the learned latent space \cite{higgins2017beta}.
FactorVAE extends $\beta$-VAE by introducing a specific objective that promotes the independence of each dimension in the latent space, which encourages the model to learn disentangled representations by explicitly maximizing the total correlation between the latent variables \cite{duan2022factorvae}. Additionally, both InfoGAN \cite{chen2016infogan} and AVB \cite{mescheder2017adversarial} are approaches that aim to enhance the disentanglement of latent representations. InfoGAN focuses on explicitly maximizing the mutual information between latent variables and the observed data, while AVB incorporates adversarial training to encourage the VAE to learn more informative and disentangled latent representations \cite{ling2022source, ling2023deep}. These methods provide alternative ways to promote disentanglement and improve the quality of generated samples in the context of generative models.

\begin{figure*}
\centering
\includegraphics[width=1\columnwidth]{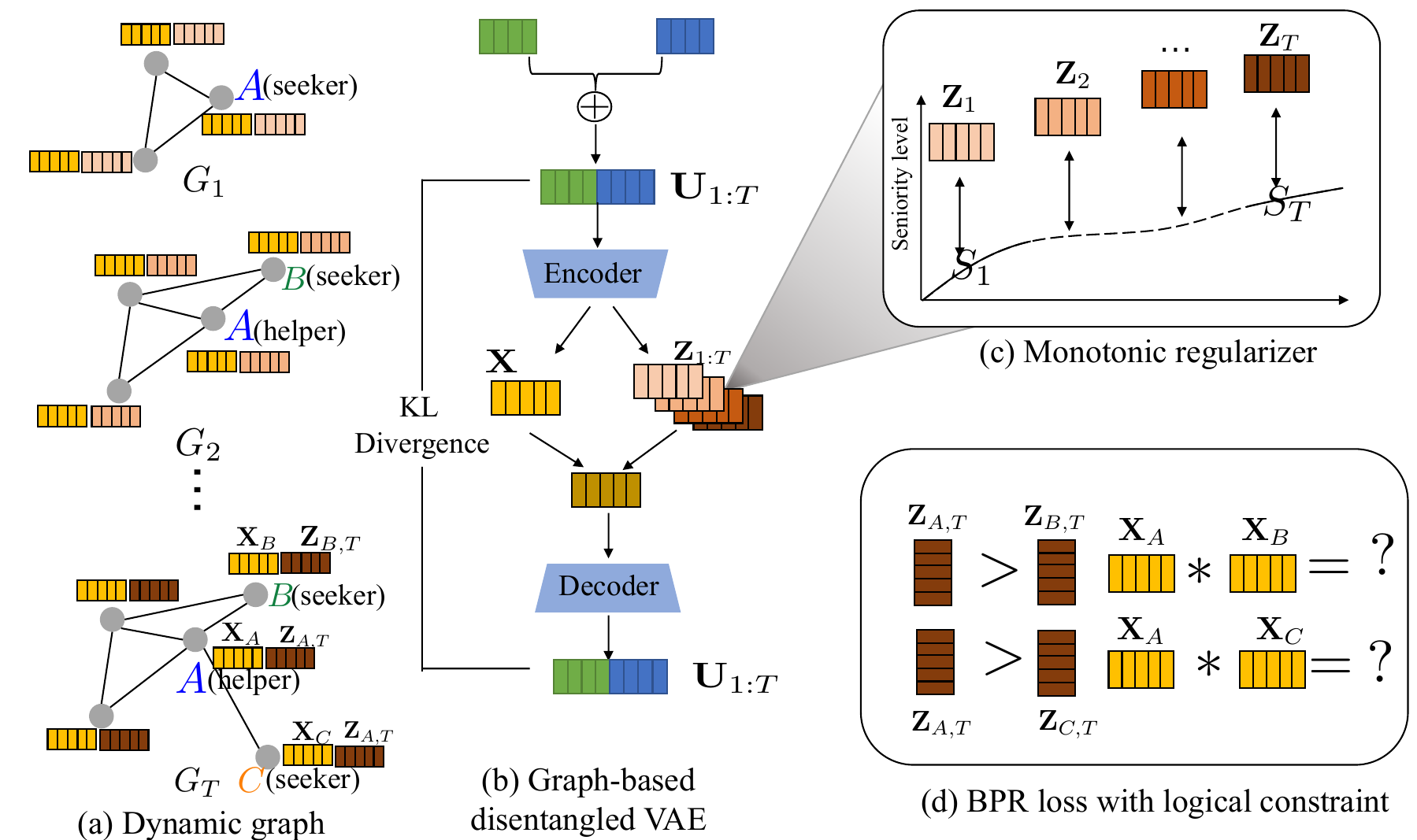}
\caption{The Framework of MINT.  (a) represents the dynamic graph that captures  the interactions between patients. Nodes denote the patients; edge denotes the helper answering the questions on particular threads at some time. (b) shows the modeling process of the historical activities of patients to obtain time-invariant and time-varying features. (c) depicts the monotonic regularizer and seniority level constraint to guarantee the logical seniority orders.}
\label{frame}
% \vspace{-0.3cm} 
\end{figure*}

\section{Proposed Method}

\subsection{Problem Formulation}
Online Health Communities (OHCs) have demonstrated their positive impact on patients by facilitating communication between answer seekers and problem helpers  \cite{young2019girl,chen2020linguistic}.
To model such communication dynamics among patients, we introduce a dynamic social support graph denoted as $G=\{G_1,\cdots,G_T\}$. Each snapshot graph $G_t=(U_t,E_t)$ captures the interactions between patients up to time $t$, as illustrated in Figure \ref{frame}(a). Here, $U_t$ represents the set of nodes (i.e., patients), $E_t$ denotes the edges formed when helpers respond to questions posed by seekers on specific threads at time $t$, and $E_t \subseteq U_t\times U_t$. $S$ denotes the set of seeker-helper pairs.
The interactions between patients are significantly influenced by heterogeneous historical activities within OHCs, such as the threads they have visited and their experienced health stages. 
Upon gaining experience, she would acquire the ability to assist patients $B$ and $C$ in resolving their issues.
To capture the influence of historical activities on patient interactions, we incorporate them as node features $\mathbf{u}_t$ at each time point $t$.
Specifically, we denote the ordered thread sequences that patients have visited as ${v}_{1:T}=({v}_1,\cdots,{v}_t,\cdots,{v}_T)$, and the health stage sequences that patients have gone through as ${h}_{1:T}=({h}_1,\cdots,{h}_t,\cdots,{h}_T)$. For instance, as depicted in Figure \ref{intro}, patient $A$ visited the thread \emph{Diagnosed}, when she was in the health stage  \emph{Just Diagnosed}. And when she progressed to the health stage \emph{Chemotherapy}, she developed an interest in the thread \emph{Intrathecal chemo}.  
The embeddings of the thread sequences and health stage sequences can be represented as $\mathbf{v}_{1:T}=(\mathbf{v}_1,\cdots,\mathbf{v}_t,\cdots,\mathbf{v}_T)$ and $\mathbf{h}_{1:T}=(\mathbf{h}_1,\cdots,\mathbf{h}_t,\cdots,\mathbf{h}_T)$, respectively. Here, $\mathbf{v}_t \in \mathbb{R}^{1\times D_v}$ denotes the embedding of the thread visited by patients at time $t$, and $\mathbf{h}_t \in \mathbb{R}^{1\times D_h}$ represents the embedding of the health stage experienced by patients at time $t$. The dimensions of the thread and health stage embeddings are denoted as $D_v$ and $D_h$, correspondingly.
Thus, at each time $t$, the embeddings of patient nodes can be computed as $\mathbf{u}_t=concat[\mathbf{v}_t,\mathbf{h}_t]$, where $\mathbf{u}_t \in \mathbb{R}^{m \times (D_v+D_h)}$. The patient node embedding sequence is denoted as $\mathbf{u}_{1:T}= (\mathbf{u}_1,\cdots,\mathbf{u}_t,\cdots,\mathbf{u}_T)$, which records the ordered historical activities of patients.
In addition, we assign a quantitative value to represent the seniority of the patient at time $t$, which means the patient's knowledge and experience of the disease, expressed as $s_t \in \mathbb{R}$ for the seniority levels of seekers and $o_t \in \mathbb{R}$ 
 for the seniority levels of helpers at time $t$. This value can be calculated based on factors such as the number of threads visited, the number of health stages experienced, and the duration of their presence on OHCs. 
Similarly, we can obtain a sequence of seniority level $s_{1:T}=(s_1,\cdots, s_t,\cdots,s_T)$ and $o_{1:T}=(o_1,\cdots, o_t,\cdots,o_T)$ corresponding to different times.

Given the visited thread sequence ${v}_{1:T}$, health stage sequence ${h}_{1:T}$, senior level sequence $s_{1:T}$, and the social support dynamic graph $G$, the objective is to recommend suitable helpers for addressing seekers' questions.
However, this problem poses significant challenges that make it highly complex to tackle: 1). Difficulty in modeling the interactions between problem helpers  and solution seekers due to heterogeneous reasons. 
 2). Difficulty in differentiating the time-invariant and time-varying influence of historical information $v_{1:t}$, $h_{1:T}$, and $s_{1:T}$.
 3). Difficulty in guaranteeing the logical seniority orders between predicted helpers and seekers. 

 \begin{table}[]
  \caption{Notations and Descriptions}
 \begin{tabular}{ll}
\toprule
Symbol & Deﬁnition\\
\midrule
$\mathbf{f}$ & latent variables\\
$\mathbf{z}_t$ & patient time-varying features\\
$\mathbf{z}_{1:T}$  & a sequence of T dynamic latent variable \\ 
% $[\mathbf{z}_t]_i$ & the i-th dimension of $\mathbf{z}_t$ \\
$\mathbf{x}$ & patient time-invariant features\\
 $s_t$, $o_t$ & senior level of seekers and helpers at time $t$\\
  ${s}_{1:T}, {o}_{1:T}$ & a sequence of T seniority level of seekers and helpers\\
${h}_{1:T}$ & a sequence of T historical health stages\\
  ${v}_{1:T}$ & a sequence of T historical threads\\
  $G_t=(U_t,E_t)$& the graph snapshot  at time $t$\\
%   \mathbf{A}& \\
\bottomrule
 \end{tabular}
 % \vspace{-3mm}
\end{table}
\subsection{The Objective of seeker-helper modeling}
To effectively address the unique recommendation problem for patients and overcome the three aforementioned challenges, we propose a novel framework for modeling seeker and helper. More concretely,  to tackle the first challenge, we introduce a novel framework to model the patient relationship based on the graph $G$. The features of nodes $U$ record the influence of patients' historical  activities, while the edges $E$ between patients signify their participation in shared threads, whether as seekers asking questions or as helpers providing answers. 
This interconnectedness allows the exchange of knowledge among connected patients, fostering a virtuous circle of mutual assistance, as depicted in Figure \ref{frame}(a).
 To overcome the second challenge, we design a graph-based Disentangled VAE to learn both time-invariant features of patients (e.g., patients' intrinsic knowledge) and time-varying features of patients (e.g., patients' evolving states across different time points)  by leveraging the modeling of historical information, including thread sequences $v_{1:t}$ and health stage sequences $h_{1:T}$. The detail is  shown in Figure \ref{frame}(b). 
%  To tackle the third challenge, we design a monotonic regularizer that enforces the patient's time-varying features $\mathbf{z}_{1:T}$  to learn the      increase monotonically patient's time-varying features $\mathbf{z}_{1:T}$   along with the seniority level $s_{1:T}$ at the corresponding time point, which can embed the seniority information into the patient's embedding during the learning process. 
 To address the third challenge, we design a monotonic regularizer that ensures the patient's time-varying feature $\mathbf{z}_{1:T}$ exhibits a monotonically increasing behavior in alignment with the seniority level $s_{1:T}$. 
 Based on the monotonically increasing properties corresponding to seniority level, we establish a seniority level constraint applied to time-varying features to characterize the roles of helper and seeker  in the prediction process. This addresses Challenge 3 and is depicted in detail in Figure \ref{frame}(c-d).

To effectively train our comprehensive model, we aim to maximize the evidence lower bound (ELBO) loss of the graph-based Disentangled VAE, denoted as $\mathcal{L}_{dis}$ (refer to Eq. \eqref{dis}), which enables the learning of both time-varying and time-invariant features. In order to capture the stability of time-varying features over short time periods, we incorporate a smoothness constraint, denoted as $\mathcal{L}_{smo}$ (refer to Eq. \eqref{smo}), which minimizes the $L_2$ distances between embeddings of adjacent time steps.  In addition to minimizing the Bayesian Personalized Ranking (BPR-loss)   $\mathcal{L}_{bpr}$ (detailed in Eq. \eqref{bpr}), which is computed based on the time-invariant features of patients, we minimize a monotonic regularization $\mathcal{L}_{reg}$ (detailed in Eq. \eqref{reg}) to enforce the time-varying features to have monotonic relation with seniority level. Moreover, in order to guarantee that the recommended helpers can have the ability to solve the seekers' problems,  a seniority-level constraint is introduced. 
 The overall objective function is shown as follows:
% \vspace{-0.1cm}
\begin{align}\label{eq:objective_function}
\min_{\Theta} \mathcal{L}&= \alpha \mathcal{L}_{dis}+\gamma\mathcal{L}_{smo}+\lambda \mathcal{L}_{bpr}+\beta \mathcal{L}_{reg},\\
\vspace{-0.1cm}
&\text{s.t.}, s_{t}<o_{t},\quad t\in [1:T]\nonumber
% \vspace{-0.2cm}
    \end{align} where $\alpha$, $\gamma$,  $\lambda$, $\beta$ are non-negative trade-off weights for each corresponding components. These weights determine the relative contribution of each component to the overall optimization process.  $\Theta$ is the set of all parameters that need to be optimized in the objective function  Eq. \eqref{eq:objective_function}.

 \subsection{Graph-based disentangled variational autoencoder\label{feature}}
The graph-based disentangled variational autoencoder (VAE) model is designed to characterize patients based on their visited threads and experienced health stages, as well as the social support graph. In this model, we assume that the sequence of patient node embeddings, denoted as $\mathbf{u}_{1:T}=(\mathbf{u}_1,\cdots,\mathbf{u}_t,\cdots,\mathbf{u}_T)$, is generated from a latent variable $\mathbf{x}$ that can be factorized into two disentangled variables: the time-invariant features $\mathbf{x}$ (representing patients' intrinsic knowledge) and the time-varying features $\mathbf{z}_{1:T}$ (representing patients' state after visiting threads at different time points). The prior distribution of the time-invariant features $\mathbf{x}$ is defined as a standard Gaussian distribution: $\mathbf{x}\sim \mathcal{N}(0,1)$. The time-varying features $\mathbf{z}_{1:T}$ follow a sequential prior, where $\mathbf{z}_t| \mathbf{z}_{<t} \sim \mathcal{N}(\mu_t,\text{diag}(\sigma_t^2))$. Here, $[\mu_t, \text{diag}(\sigma_t^2)]=f_\theta(\mathbf{z}_{<t} )$, and $\mu_t$ and $\text{diag}(\sigma_t^2)$ are the parameters of the prior distribution conditioned on all previous time-varying features $\mathbf{z}_{<t}$. The model parameter $\theta$ can be parameterized as a recurrent network, such as LSTM or GRU, where the hidden state is updated over time.
% Overall, the prior distribution of the latent variable $\mathbf{f}$ can be factorized as follows:
% \begin{equation*}
% p(\mathbf{f})=p\left(\mathbf{x}|G_{T}\right) p\left(\mathbf{z}_{1: T}\right)=p\left(\mathbf{x}|G_{T}\right) \prod_{t=1}^{T} p\left(\mathbf{z}_{t} \mid \mathbf{z}_{<t}\right)
% % \vspace{-0.3cm}
% \end{equation*}
We define the generating distribution of patient features conditioned on the time-invariant features $\mathbf{x}$ and time-varying features $\mathbf{z}_t$ as follows:
\begin{align*}
p(\mathbf{u}_{1:T},\mathbf{z}_{1:T}, \mathbf{x} ,G_{1:T})&=p(\mathbf{x})\prod_{t=1}^{T} p\left(\mathbf{u}_{t} \mid \mathbf{x}, \mathbf{z}_{t},G_{t}\right)p(\mathbf{z}_{t}|\mathbf{z}_{<t})\\
&\mathbf{u}_t| \mathbf{z}_{t},\mathbf{x},G_{t} \sim \mathcal{N}(\mu_{u,t},\text{diag}(\sigma_{u,t}^2))
\end{align*}
where $[\mu_{u,t},\text{diag}(\sigma_{u,t}^2)]=g_\phi( \mathbf{z}_{t},\mathbf{x} |G_{t})$.
During the inference process, the graph-based Variational Autoencoder  employs variational inference to approximate the posterior distributions. Specifically, we approximate the posterior distributions as follows:
% \vspace{-0.cm}
\begin{align*}
&\mathbf{x}|G_{T} \sim \mathcal{N}\left(\mathbf{\mu}_{x}, \operatorname{diag}\left(\sigma_{x}^{2}\right)\right), \quad \mathbf{z}_{t} \sim \mathcal{N}\left(\mathbf{\mu}_{t}, \operatorname{diag}\left(\sigma_{t}^{2}\right)\right)\\
&[\mu_{x},\text{diag}(\sigma_{x}^2)]=h_\delta( \mathbf{u}_{1:T}),[\mu_{u,t},\text{diag}(\sigma_{u,t}^2)]=h_\eta( \mathbf{u}_{\leq{t}})
\end{align*}
where the parameters $\delta$ and $\eta$ correspond to the recurrent encoder functions $h_\delta$ and $h_\eta$, respectively, which are responsible for computing the approximated posterior distributions. Specifically, the time-invariant features are conditioned on the entire sequence and encoded using $h_\delta$, whereas the time-varying features are inferred using $h_\eta$ and conditioned only on the previous time point. Thus, our inference model can be factorized as follows:
% \vspace{-0.1cm}
\begin{equation*}
q\left(\mathbf{z}_{1: T}, \mathbf{x} \mid \mathbf{u}_{1: T},G_{1:T}\right)=q\left(\mathbf{x} \mid \mathbf{u}_{1: T},G_{1:T}\right) \prod_{t=1}^{T} q\left(\mathbf{z}_{t} \mid \mathbf{u}_{\leq t}\right)
\end{equation*}
The loss of graph-based disentangled VAE in \eqref{eq:objective_function} is a timestep-wise negative variational lower bound:
% \begin{equation}
% \begin{aligned}
% &\mathcal{L}^{(p)}_{V A E} =\mathbb{E}_{q\left(\mathbf{z}^{(p)}_{1: T}, \mathbf{x}^{(p)} \mid \mathbf{u}^{(p)}_{1: T}\right)}\left[-\sum_{t=1}^{T} \log p\left(\mathbf{u}^{(p)}_{t} \mid \mathbf{x}^{(p)}, \mathbf{z}^{(p)}_{t}\right)\right]+\\
% & \mathrm{KL}\left(q\left(\mathbf{x}^{(p)}\mid \mathbf{u}^{(p)}_{1: T}\right) \| p\left(\mathbf{x}^{(p)}\right)\right)+\sum_{t=1}^{T} \mathrm{KL}\left(q\left(\mathbf{z}^{(p)}_{t}\mid \mathbf{u}^{(p)}_{\leq t}\right) \| p\left(\mathbf{z}^{(p)}_{t} \mid \mathbf{z}^{(p)}_{<t}\right)\right)
% \end{aligned}
% \end{equation}
% \begin{equation}
% \begin{aligned}
% &\mathcal{L}_{d i s} =\mathbb{E}_{q\left(\mathbf{z}_{1: T}, \mathbf{x} \mid \mathbf{u}_{1: T},G_{T}\right)}\left[-\sum_{t=1}^{T} \log p\left(\mathbf{u}_{t} \mid \mathbf{z}_{t},\mathbf{x}, G_{T}\right)\right]+\\
% & \mathrm{KL}\left(q\left(\mathbf{x}\mid \mathbf{u}_{1: T},G_{T}\right) \| p\left(\mathbf{x}|G_{T}\right)\right)+\sum_{t=1}^{T} \mathrm{KL}\left(q\left(\mathbf{z}_{t}\mid \mathbf{u}_{\leq t}\right) \| p\left(\mathbf{z}_{t} \mid \mathbf{z}_{<t}\right)\right)
% \end{aligned}
% \label{dis}
% \end{equation}
\begin{equation}
\begin{aligned}
\mathcal{L}_{dis } & =\mathbb{E}_{q\left(\mathbf{z}_{1: T}, \mathbf{x} \mid \mathbf{u}_{1: T}, G_{1:T}\right)}\left[-\sum^{T}_{t=1} \log p\left(\mathbf{u}_{t} \mid \mathbf{z}_{t}, \mathbf{x}, G_{t}\right)\right] \\
& +\mathrm{KL}\left(q\left(\mathbf{x} \mid \mathbf{u}_{1: T}, G_{1:T}\right) \| p\left(\mathbf{x} \mid G_{1:T}\right)\right) \\
& +\sum_{t=1}^{T} \mathrm{KL}\left(q\left(\mathbf{z}_{t} \mid \mathbf{u}_{\leq t}\right) \| p\left(\mathbf{z}_{t} \| p\left(\mathbf{z}_{t} \mid \mathbf{z}_{<t}\right)\right)\right.
\end{aligned}
\label{dis}
\end{equation}
Additionally, the smoothing constraint in Equation \eqref{eq:objective_function} can be written as follows:
\begin{equation}
\mathcal{L}_{smo}=\|\mathbf{z}_{t}-\mathbf{z}_{t-1}\|_2^2
\label{smo}
\end{equation}
We capture the interactions between patients in terms of question-answering dynamics within the social support graph. The adjacency matrix of the social support graph is defined as follows:
\begin{equation*}
\mathcal{A}_T=\left(\begin{array}{cc}
0 & \mathcal{R}_T \\
\mathcal{R}_T^\mathrm{T} & 0
\end{array}\right)
\end{equation*}the interaction matrix $\mathcal{R}_T \in \mathbb{R}^{m \times m}$ represents the connections between patients within the social support graph. Each entry of $\mathcal{R}$ is assigned a value of 1 if there is an interaction between the corresponding pair of patients, otherwise, it is 0.
To initialize the patient nodes on the graph, we use the obtained time-invariant features $\mathbf{x}$ as the embeddings for the 0-th layer, denoted as $\mathbf{e}^{(0)}=\mathbf{x}$.
Subsequently, the patient embeddings at the $l$-th layer can be expressed as:
% \vspace{-0.3cm}
\begin{equation*}
\mathbf{e}^{(l)}=\left(\mathcal{D}^{-\frac{1}{2}} \mathcal{A} \mathcal{D}^{-\frac{1}{2}}\right) \mathbf{e}^{(l-1)}
% \vspace{-0.2cm}
\end{equation*}where $\mathcal{D}^{-\frac{1}{2}} \mathcal{A} \mathcal{D}^{-\frac{1}{2}} $ is used to compute the symmetrically normalized matrix.  $\mathcal{D}$ represents a $2m \times 2m$ degree matrix. Each entry $\mathcal{D}_{ii}$ in $\mathcal{D}$ corresponds to the number of nonzero entries in the i-th row vector of the adjacency matrix $\mathcal{A}$.  As a result, the final time-invariant features of patients can be represented as:
% \vspace{-0.2cm}
\begin{equation*}
\mathbf{e}=\frac{1}{L}\sum_{l=0}^{l=L} \mathbf{e}^{(l)}
% \vspace{-0.2cm}
\end{equation*}

 \subsection{ Recommending under logical seniority orders\label{recommend}}
The learned patient  embeddings $\mathbf{e}$ capture both the intrinsic time-invariant information and collaborative signals.
To differentiate between seekers and helpers, we utilize $\mathbf{e}_p \in \mathbb{R}^{a \times D_p}$ and $\mathbf{e}_q \in \mathbb{R}^{b \times D_q}$, where $a\leq m$, $b \leq m$, $D_p$, and $D_q$ represent the dimensionality of the time-invariant features.
Moreover, the roles of seekers and helpers are determined by their activities, specifically whether they answer or ask questions on the threads within OHCs. The BPR-loss in Eq. \eqref{eq:objective_function} can be expressed as:
\begin{equation}
\mathcal{L}_{bpr}=-\sum_{(a,+,-)\in S}\ln \sigma \left(\hat{r}_{a,+}-\hat{r}_{a,-}\right) \\
\label{bpr}
% \vspace{-0.2cm}
\end{equation}
where $\hat{r}_{a,+}=\mathbf{e}_{p}^{(a)}{\mathbf{e}_{q}^{(+)}}^T$ calculates the similarity score  between a seeker and a positive sampled helper,
$\hat{r}_{a,-}=\mathbf{e}_{p}^{(a)}{\mathbf{e}_{q}^{(-)}}^T$ computes the similarity between seekers and negatively sampled patients.
However, while the BPR-loss can learn the similarities between seekers and helpers, it does not guarantee that the recommended helpers possess sufficient experience to assist the seekers. In order to address this, we introduce a monotonic regularizer to enforce a monotonic correlation between the time-varying features $\mathbf{z}_t$ and the seniority level.
To derive the regularizer, we refer to the standard definition of monotonicity \cite{pemberton2007mathematics}: if $f(x)\geq f(y)$, then $x \geq y$, where $x$ and $y$ represent any latent variables. However, directly computing such a non-linear relationship is challenging. Instead, we penalize the following term in the objective as an equivalent formulation: $\max \left(0,-\left(f(x)-f(y)\right) \cdot\left(x-y\right)\right)$. Take seekers as an example, in our specific context, the monotonic correlation is implemented through the regularizer term $\mathcal{L}_{reg}$, which is defined as:
% \vspace{-0.2cm}
 \begin{equation}
\mathcal{L}_{reg}=\sum^n_i\text{ReLU}[ (s_t-s_{t-1})*([\mathbf{z}_{t-1}]_i-[\mathbf{z}_{t}]_i)]
\label{reg}
% \vspace{-0.2cm}
\end{equation}where $n$ represents the embedding size of time-varying features and $i$ denotes each entry of the embeddings, ReLU \cite{glorot2011deep} is  the activation function. For helpers, $s$ can be replaced by $o$. The regularizer, which ensures the monotonic correlation, can be incorporated into the feature learning loss given in Eq. \eqref{dis}. This addition strengthens the training objective by encouraging the desired monotonic behavior in the time-varying features.
Building upon the monotonically regularized time-varying features, we further introduce a seniority-level constraint to enforce the logical ordering between the recommended helpers and seekers. That is, at the same time $t$, each dimension of the helpers must be larger than each dimension of the corresponding seekers. This constraint between the pairs seeker $p$ and the helper $q$ is formulated as:
% \vspace{-0.2cm}
\begin{equation}
s.t., [\mathbf{z}_{p,t}]_i<[\mathbf{z}_{q,t}]_i
% \vspace{-0.2cm}
\end{equation}
where $[\mathbf{z}_{p,t}]_i$,  $[\mathbf{z}_{q,t}]_i$  denote the i-th entry of the time-varying features for seeker  and helper, respectively. A computable form of this constraint can be written as:
\begin{equation}
    \mathcal{L}_{cons} =  \sum_{\{p,q\} \in S} \sum_i ([\mathbf{z}_{p,t}]_i-[\mathbf{z}_{q,t}]_i)
\end{equation}

We put $\mathcal{L}_{cons}$ into $\mathcal{L}_{reg}$, which is used to train the model together.
By enforcing the time-varying features to monotonically increase with the seniority level, this constraint ensures that the recommended helpers possess the necessary capabilities to address the problems of the seekers.
% We  add the constraint to the Equation \eqref{bpr}. 

\section{Experiments}

Extensive experiments were conducted to assess the performance of our proposed method, MINT, on two real-world Online Health Community (OHC) datasets. To establish the effectiveness of MINT, we compared it with four other methods. Furthermore, we analyzed the trade-off between efficiency and accuracy, providing insights into the balance achieved by our approach. Sensitivity analysis was also performed to examine the impact of hyperparameters in MINT. Ablation studies were conducted to underscore the significance of the designed components within the Helper-Seeker recommendation framework. Lastly, a case study was conducted, highlighting the necessity of the recommendation for seekers in OHCs. All experiments were conducted with a single Tesla V100 GPU.

\subsection{Dataset and Data preparation}
The Breast Cancer Community stands as one of the largest Online Health Communities (OHCs), offering a platform for patients to acquire knowledge about the disease, seek and provide social support, and connect with others sharing similar health stages, as evident in the signatures of their accounts  \cite{gao2022modeling}. Interactions among patients within this community are recorded in various threads, comprising questions and corresponding answers. Furthermore, patients navigate through different threads over time, reflecting their evolving health stages. The dataset spans from the beginning of 2014 to the end of 2018. Similarly, the Bladder Cancer Community serves as a popular OHC where patients can exchange bladder cancer-related information. The dataset utilized in this study covers the period from 2006 to 2021 and shares the same information characteristics as the Breast Cancer Community dataset. In both datasets, we filtered out threads with fewer than ten patient visits and disregarded non-helping interactions. Patients' visited threads were organized chronologically, along with the interactions occurring within those threads.
We also filtered out irrelevant and unhelpful replies and identify helper and seeker roles by differentiating patient activities (i.e. asking and answering in threads, including professional help and emotional support, etc.). It is worth noting that the role of the patient is not fixed, early cancer patients are more inclined to ask questions, at this time they are seekers, as time goes on, they visit more threads or have recovered, and they are usually helpers to answer other patients' questions.
Each dataset was divided into three subsets: a training set, a validation set, and a testing set, with an 80\%, 10\%, and 10\% ratio, respectively. Further details regarding the dataset statistics can be found in Table \ref{tab:data}.

\begin{table}[t]
  \centering
% \footnotesize
 \caption{Analysis of two OHC datasets.}
  \begin{tabular}{lrr}
    \toprule
    &Breast Cancer OHC& Bladder Cancer OHC\\
    \midrule
    \#Patient & 3,948&296\\
\#Seeker & 719&189 \\
   \#Helper&3,827&243 \\
   \#Interaction& 16,360&9,867\\
%     Ratio &0.5493 & 0.5109\\
%     sub-community Density& 2.430 &5.215\\
%     thread Density&0.418&0.017\\
  \bottomrule
\end{tabular}
% \vspace{-0.3cm}
  \label{tab:data}
% \vspace{-0.3cm}
\end{table}

\subsection{Experiment Setup}
\subsubsection{Comparison methods.} In order to demonstrate the effectiveness of our proposed MINT method, we conducted comparisons with state-of-the-art methods from three distinct categories. 1) \emph{Traditional method:} BPR-MF \cite{rendle2009bpr} is a well-established recommendation algorithm known for its strong performance across various recommendation tasks. 2) \emph{Sequential methods:}  
SASRec \cite{kang2018self}  has shown promising performance in sequential recommendation tasks, particularly in scenarios where temporal dependencies and complex patterns are crucial. 
 3) \emph{Graph-based methods:} 
NGCF \cite{wang2019neural} and LightGCN \cite{lightgcn} leverage the bipartite graph structure of user-item interactions to learn user and item embeddings, facilitating effective recommendations.  DGCF  \cite{wang2020disentangled}  seeks to overcome limitations in traditional methods by explicitly modeling and disentangling different aspects of user preferences and item characteristics, but ignores the temporal information.
GraFRank \cite{sankar2021graph} stands as the pioneering work in exploring the utilization of graph neural networks for modeling social user-user interactions and recommendations. It effectively captures expressive user representations through the integration of multiple feature modalities and user-user interactions. 
% By including these representative methods from each category, we aimed to provide comprehensive comparisons and demonstrate the superiority of our proposed MINT method.
% \\
% \vspace{-0.4cm}
% \begin{itemize}[leftmargin=*]
% \item \textbf{BPR-MF}   is a conventional  recommendation algorithm and can provide top-k recommendation, which uses MF as a model with the learning objective of BPR.
% \item \textbf{NGCF}   models user-item interactions on a bipartite graph to learn the embeddings of users and items using GCN.
% \item \textbf{LightGCN}    takes the advantages of neighborhood aggregation to characterize  collaborative filtering information  without feature transformation and nonlinear activation.
% \item \textbf{GraFRank} proposes a neural architecture based on graph to learn expressive user representations from multiple feature modalities and user-user interactions.
% \vspace{-0.1cm}
% \end{itemize}
% \textbf{BPR-MF}   is a conventional  recommendation algorithm and can provide top-k recommendation, which uses MF as a model with the learning objective of BPR.\\
% \textbf{NGCF}   models user-item interactions on a user-item bipartite graph to learn the embeddings of users and items using GCN.\\
% \textbf{LightGCN}    takes the advantages of neighborhood aggregation to characterize  collaborative filtering information  without feature transformation and nonlinear activation.\\
% \textbf{GraFRank} proposes a neural architecture based on graph to learn expressive user representations from multiple feature modalities and user-user interactions.
\subsubsection{Implementation Details.}
For the proposed model, we utilized two LSTMs to encode the time-varying features, while one LSTM combined with a one-layer perception was employed for encoding the time-invariant features. The decoder comprised two-layer perceptions with non-linear transformations. 
To obtain patient embeddings, we concatenated the visited thread and health stage embeddings, resulting in 16 dimensions. The dimensions of the time-varying features $\mathbf{z}_t$ and time-invariant features $\mathbf{x}$ were both set to 8. We found that setting the graph layer to 3 yielded the best performance.
During model training, we utilized a batch size of 256 and employed the Adam optimizer with a learning rate of 0.001. The hyperparameter $\lambda$ for the bpr-loss was set to 1, and the smoothing coefficient $\gamma$ was set to 0.1. The effect of parameters $\alpha$ and $\beta$ is reported in Section \ref{hyper}.

In contrast, the original BPR-MF, NGCF, DGCF, LightGCN, and GrafRank methods solely focused on collaborative filtering signals between entities, disregarding the historical visiting activities of patients. To ensure fair comparisons, we incorporated LSTM to encode the historical patient information, using the resulting embeddings as the initial feature representations to capture personalized information.
In SASRec, the
max sequence length was set to 10, which is roughly proportional to the mean number of interactions per patient.
Similar to our model, the embedding dimensions were set to 16. We employed a batch size of 256 and used the Adam optimizer with a learning rate of 0.001 for training. The layer parameter for graph-based models was also set to 3, as indicated in the corresponding papers. The factors of disentanglement for DGCF were set to 4.

\subsubsection{Evaluation Metrics.}
The evaluation of each method's performance is conducted using established metrics, including the Mean Reciprocal Rank (MRR) \cite{voorhees1999trec}, Normalized Discounted Cumulative Gain (NDCG@K) \cite{jarvelin2002cumulated}, and HIT@K \cite{powers2011evaluation}. These metrics are widely used to assess the quality of recommendations, with higher scores indicating superior recommendation performance. Specifically, the MRR measures the rank of the ground truth entity relative to all other entities. NDCG@K evaluates the relevance of recommended entities based on their positions in the ranking list, considering the discounted cumulative gain. Lastly, HIT@K indicates the proportion of correct recommendations within  top K positions.

\begin{table*}[h]
    % \vspace{-0.2cm}
  \centering
  \begin{threeparttable}
  \caption{Experimental results on Breast Cancer Community dataset with respect to four evaluation metrics. The improvement of the proposed method over baselines is significant at level $\alpha$ of 5\%. }
  \label{result1}
    \begin{tabular}{cccccccc}
    \toprule
  Method &
    \multicolumn{6}{c}{ Breast Cancer Community Dataset }\cr
    \cmidrule(lr){2-8} 
    & NDCG@3 & HIT@3 & NDCG@5 & HIT@5 & NDCG@10 & HIT@10 &MRR\cr
    \midrule
    BPR-MF &   0.0151    &   0.0204   &  0.0177   &  0.0267     & 0.0220     &   0.0402   &0.0215    \cr
     SASRec & 0.0144 &   0.0149 &   0.0162   & 0.0194   &  0.0223   &  0.0388  &  0.0241    \cr
   NGCF &  0.0096     &  0.0124     &   0.0113   &    0.0165   &  0.0145     &    0.0264 & 0.0144  \cr
LightGCN  &    0.0139    &   0.0177    &    0.0167    &  0.0245    &     0.0225    &    0.0425   & 0.0220   \cr
  DGCF &  0.0132   &   0.0122   &  0.0152    &   0.0138   & 0.0201  & 0.0366 &   0.0149   \cr
GraFRank &  \underline{0.0187}     &    \underline{0.0248}   &  \underline{0.0216}      &   \underline{0.0321}    &     \underline{0.0267}    &  \underline{0.0458}   &   \underline{0.0258}  \cr
Ours     &   \textbf{0.0244}   &   \textbf{0.0281}    &    \textbf{0.0270}    &   \textbf{0.0345}    &    \textbf{0.0327}    &    \textbf{0.0526}    & \textbf{0.0278}\cr
  \bottomrule
    \end{tabular}
    \end{threeparttable}
\end{table*}

\begin{table*}[h]
    % \vspace{-0.2cm}
  \centering
  \begin{threeparttable}
  \caption{Experimental results on Bladder  Cancer Community dataset with respect to four evaluation metrics. The improvement of the proposed method over baselines is significant at level $\alpha$ of 5\%. }
  \label{result2}
    \begin{tabular}{cccccccc}
    \toprule
  Method &
    \multicolumn{6}{c}{ Bladder  Cancer Community Dataset }\cr
    \cmidrule(lr){2-8} 
    & NDCG@3 & HIT@3 & NDCG@5 & HIT@5 & NDCG@10 & HIT@10 &MRR\cr
    \midrule
   PR-MF   &    0.1672   &   0.2110   &   0.1972     &  0.2836  &   0.2345  &   0.3991 &0.1998\\
    SASRec &    0.2156  &    0.2604 &  0.2304     &   0.2959    &   0.2400    &   0.3254  &   0.2217  \cr
NGCF   &   0.1809   &   0.1298 &  0.3212    & 0.1862   &  0.5274      &  0.2542 &0.1820\\
LightGCN &     0.2556   &   0.3249    &    0.2941     &   0.4200    &     0.3295     &    0.5272   &  0.2781 \\
 DGCF &    0.3511   &   0.2960     &   0.4383     &   0.4240     &    0.5652    & 0.6329  &  0.4418  \cr
GraFRank &\underline{0.4618} &   \textbf{0.5522}  &   \underline{0.4380 }    & \underline{0.5881}    & \underline{0.4851}      &   \underline{0.6135}&   \underline{0.4467}\\
Ours     &    \textbf{0.4861}    &   \underline{0.5396}    &    \textbf{0.5120}    &   \textbf{0.6029}    &   \textbf{0.5505}     &   \textbf{0.7194}     & \textbf{0.5071}
\cr
  \bottomrule
    \end{tabular}
    \end{threeparttable}
\end{table*}

\subsection{Results}
The comprehensive performance evaluation of both the compared methods and the proposed method, MINT, is presented in Table \ref{result1} for the Breast Cancer Community dataset and Table \ref{result2} for the Bladder Cancer Community dataset. The evaluation metrics used include NDCG@3, HIT@3, NDCG@5, HIT@5, NDCG@10, HIT@10, and MRR. Each table contains results obtained from separate experiments, and higher values indicate better performance. The best-performing method is indicated in bold, while the second best-performing method is underlined. During the validation phase, MINT achieved the highest performance on the Breast Cancer Community dataset when the parameters $\alpha$ and $\beta$ were set to 0.01 and 0.001, respectively. For the Bladder Cancer Community dataset, the best performance was obtained with parameter values of $\alpha$ = 0.001 and $\beta$ = 0.01.  Statistical analysis reveals that the performance improvement achieved by the proposed method over each comparison method is statistically significant at a significance level of 5\%, except for the prediction results on the online Bladder Cancer Community dataset, where HIT@3 does not show significant improvement.

MINT demonstrates superior performance compared to the other comparison methods across various evaluation metrics. Notably, among the different types of comparison methods, NGCF, a graph-based approach, exhibits the least competitive performance. This can be attributed to the potential degradation of prediction performance caused by non-linear activation functions and feature transformation matrices.
BPR-MF performs worse than LightGCN, indicating that the incorporation of high-order information aggregation and propagation can enhance recommendation performance. 
Compared with BPR-MF, SASRec can achieve relatively better results, because the characteristics of time serialization are considered. However, because it does not consider the time characteristics of different angles and does not have the advantages of a full graph neural network, the effect is worse than MINT. Given that DGCF and GrafRank model features from different angles based on known information, they can achieve better performance than the other comparison methods.
Meanwhile, GraFRank, which incorporates attention mechanisms to consider relations between different modeled content such as experienced health stages and visited threads, achieves the best performance among the comparison algorithms.
Our proposed method, MINT, achieves highly competitive performance on both datasets. Particularly, MINT outperforms the comparison approaches with relative gains of 17.33\% and 9.52\% across each evaluation metric on the Breast Cancer Community dataset and the Bladder Cancer Community dataset, respectively.
MINT consistently outperforms the classic methods, BPR-MF, NGCF, and LightGCN, due to the complexity of the seeker-helper recommendation task in OHCs, which involves heterogeneous historical information about patients. In contrast, the classic methods fail to effectively explore the influence of historical information on patient interactions.
While GraFRank exhibits slightly better performance than MINT, its overall prediction performances still fall short in comparison to our proposed method.

\begin{table}[h]
\caption{Ablation Study on the two OHC datasets.}
    \centering
 % \footnotesize
    \begin{threeparttable}   
      \begin{tabular}{ccccccccc}
%  \small
      \toprule
    Method &
      \multicolumn{4}{c}{ Breast Cancer Community}\cr
      \cmidrule(lr){2-5}&NDCG@5&HIT@5&NDCG@10&HIT@10& \cr
      \midrule
      w/ VAE&0.0192&    0.0274        &0.0248   &0.0449          \cr
      w/o Senior&0.0196&0.0275             &0.0253    &0.0429            \cr
    Ours&0.0270&0.0345        &0.0327    &0.0526          \cr
     \toprule
      Method &
        \multicolumn{4}{c}{ Bladder Cancer Community}\cr
        \cmidrule(lr){2-5}&NDCG@5&HIT@5&NDCG@10&HIT@10& \cr
        \midrule
        w/ VAE&0.4971     &0.5753        &0.5301     &0.6735        \cr
        w/o Senior&0.5012    &0.5793        &0.5331   &0.6747            \cr
      Ours&0.5120  &0.6029  &0.5505   &0.7194  \cr
        \bottomrule
      \end{tabular}
      \end{threeparttable}
  \label{ablation1}
% \vspace{-0.4cm} 
  \end{table}

%   \begin{table}[h]
%     %   \centering
%   \label{ablation2}
%       \begin{threeparttable}
%       \scalebox{0.8}{
%         \begin{tabular}{ccccccccc}
%         \toprule
%       Method &
%         \multicolumn{4}{c}{ Bladder Cancer Community}\cr
%         \cmidrule(lr){2-5}&NDCG@5&HIT@5&NDCG@10&HIT@10& \cr
%         \midrule
%         w/ VAE&0.4971     &0.5753        &0.5301     &0.6735        \cr
%         w/o Senior&0.5012    &0.5793        &0.5331   &0.6747            \cr
%       Ours&0.5120  &0.6029  &0.5505   &0.7194  \cr
%         \bottomrule
%         \end{tabular}}
%         \end{threeparttable}
%      \caption{Ablation Study on Bladder Cancer Community data}
%     \end{table}
\subsection{Ablation Study}

    We further conduct the ablation study to investigate the importance of each component of MINT.
    We present two variants of the proposed MINT: (1) For the first ablated model, instead of modeling the time-invariant and time-varying features of patients, we only learn the time-varying features with VAE and apply monotonic regularization, noted as w/ VAE.  (2)   For the second ablated model,  we do not use the monotonic regularizer and seniority level constraint to enforce the learning process of time-varying features, noted as w/o Senior. 
Table.\ref{ablation1} represents NDCG@5, HIT@5, NDCG@10, HIT@10  of the three models over the Breast Cancer Community and Bladder Cancer Community. Overall,  we can observe that the performance will degrade if any components of our proposed MINT are removed. Only leveraging VAE without modeling the time-invariant and time-varying features can not learn the intrinsic knowledge of patients which  can be used to calculate the similarity between patients.
Additionally,  w/o Senior  has already achieved comparable performance on the Bladder Cancer Community dataset compared with other approaches in Table.\ref{result2}. However,  without a monotonic regularizer and seniority level constraint, we can only compute the similarity between patients, making it challenging to ensure that the recommended patient has the ability to answer the question. Thus, the performance of w/o Senior consistently underperforms the default settings.  
\begin{figure} [h]
\centering
\includegraphics[width=\columnwidth]{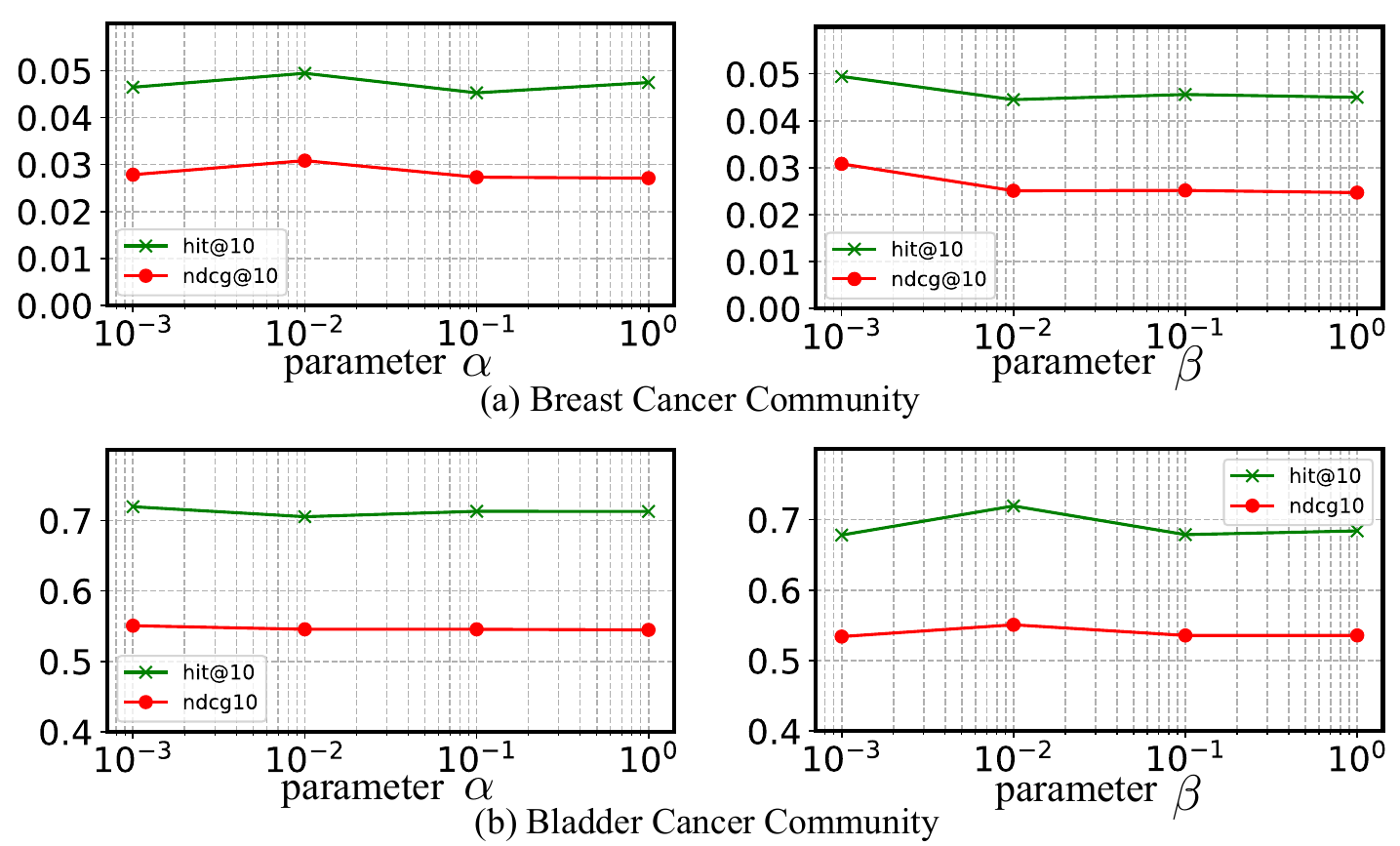}
\caption{The sensitivity of parameter $\alpha$ and $\beta$ for MINT. When tuning $\alpha$,  $\beta$ was fixed to 0.001 for Breast Cancer OHC and 0.01 for Bladder Cancer OHC. When tuning $\alpha$,  $\beta$ was fixed to 0.01 for Breast Cancer OHC and 0.001 for Bladder Cancer OHC.}
\label{para}
\vspace{-0.3cm} 
\end{figure}

\subsection{Impact of Hyper-parameters\label{hyper}}
%Next, it is crucial to investigate how parameter settings affect performance. In this section, we explore two factors: the coefficient parameter $\alpha$ and $\beta$, which were set as 0.001, 0.01, 0.1, 1. The impacts on NDCG@10 and HIT@10 are shown in Figure \ref{para}. Generally,  we can observe MINT is not sensitive to the two parameters.
%For Breast Cancer Community dataset, when tuning $\alpha$, we fixed $\beta$ to 0.001. MINT can achieve the best performance when $\alpha$ was set as 0.01. Similarly, when tuning $\beta$, we fixed $\alpha$ to 0.01, and MINT performs best when $\beta$ was set as 0.001.
%For Bladder Cancer Community dataset, when $\alpha$ and $\beta$ were set as 0.001 and 0.01 separately, MINT shows the best performance.
%Combined with ablation study in Table \ref{ablation1}  for analysis, we can imply  although the BPR-Loss is significant for learning,
%both disentangled feature learning and seniority level constraint can help correct the learning process. They can  avoid the MINT only learns the similarity information, ignoring the seniority orders between seekers and predicted helpers, which is more meaningful to real-world applications.
In the following parameter analysis, we examine the sensitivity of two important coefficient parameters, namely $\alpha$ and $\beta$, and investigate their impact on the model performance.  By fixing one parameter and tuning the other, we evaluate the performance in terms of NDCG@10 and HIT@10 under different parameter settings. The results are visualized in Figure \ref{para}. For the Breast Cancer Community dataset, we initially set $\beta$ to 0.001 and $\alpha$ to 0.01 and then varied the other parameter from the range [0.001, 0.01, 0.1, 1]. In Figure \ref{para} (a), we observe consistent performance without significant fluctuations, indicating the robustness of the proposed method to parameter changes. Similarly, for the Bladder Cancer Community dataset, MINT demonstrates superior performance when $\alpha$ and $\beta$ are individually set to 0.001 and 0.01.
When we consider the findings from the ablation study presented in Table \ref{ablation1}, it becomes apparent that both disentangled feature learning and seniority-level constraints play crucial roles in improving the learning process. Although BPR-Loss is essential for learning, it is the incorporation of these additional components that prevents MINT from solely focusing on similarity information and disregarding the hierarchical seniority orders between seekers and predicted helpers. This aspect is particularly valuable for real-world applications.
Overall, the parameter analysis and ablation study emphasize the significance of disentangled feature learning and seniority-level constraints in enhancing the performance and meaningfulness of the MINT approach.

\subsection{Case Study}
\begin{figure}
\centering
\includegraphics[width=0.7\columnwidth]{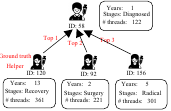}
\caption{The case study of MINT.  The top 3 predicted results for patient ID 58 in the Bladder Cancer Community are given as an example. `Years' represents the length of time that the patient came to the OHC, `Stages' and `\#thread' respectively represent the health stage when asking or answering questions, and the number of threads visited.}
\label{case}
% \vspace{-0.3cm} 
\end{figure}

\begin{figure}
\centering
\includegraphics[width=\columnwidth]{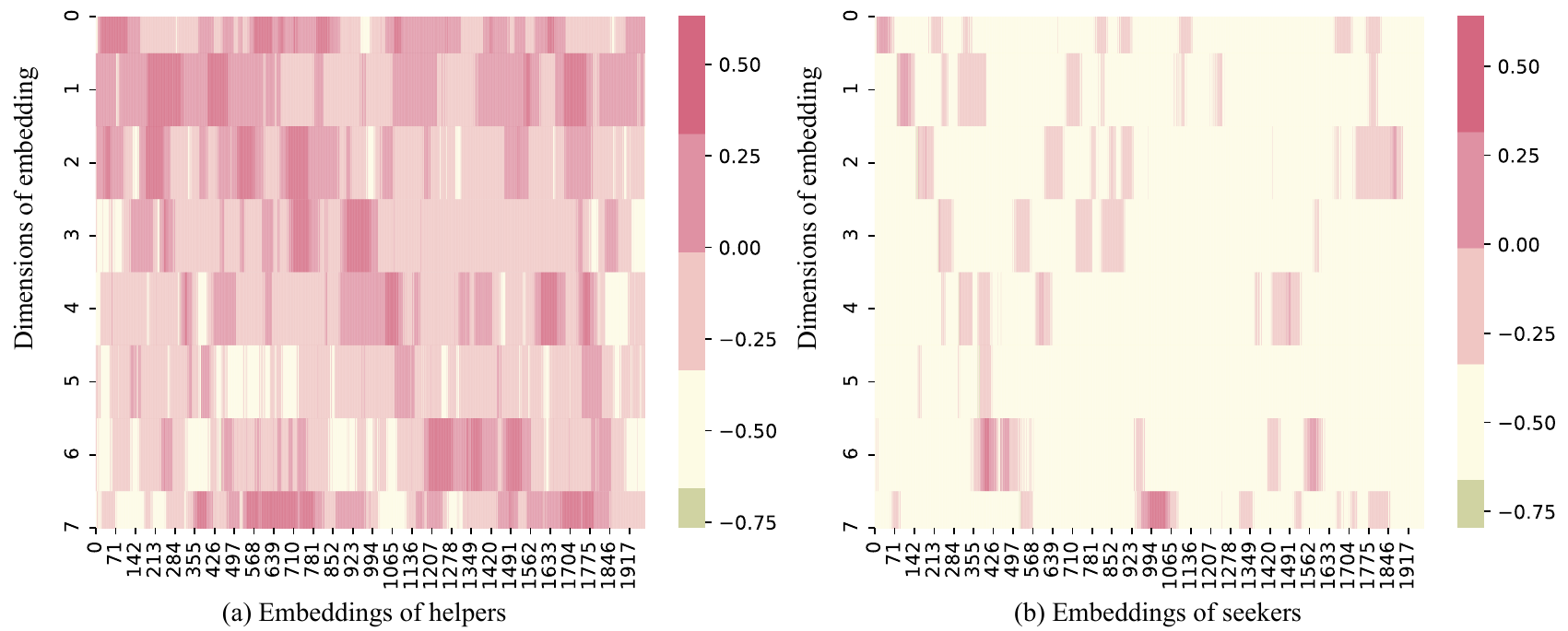}
\caption{The visualization of the seeker and helper representations.}
\label{case2}
\vspace{-0.5cm} 
\end{figure}

In addition to its performance effectiveness, MINT also ensures that the recommended helpers possess the capability to provide solutions to the seekers. Figure \ref{case} illustrates a case study to exemplify this aspect. The patient under consideration (ID: 58) joined the OHC just a year ago and is currently in the health stage of \emph{Diagnosed}. This patient has actively visited 122 threads seeking assistance. By employing the proposed MINT, we predict the top K=3 potential helpers.
From the visualization, it is evident that MINT successfully identifies the ground truth helper (patient ID: 120) who possesses the necessary knowledge to address the questions raised by patient ID: 58. This prediction is based on several factors:
Patient ID: 120 has a significantly longer tenure in the forum (joined 13 years ago) and has already overcome the illness (currently under the stage of \emph{Recover}).
Compared to other recommended patients, patient ID: 120 exhibits a combination of these crucial factors, making them the most suitable helper with the highest seniority.
Therefore, the case study demonstrates that MINT's recommendation mechanism takes into account relevant factors such as experience and seniority, ensuring the selection of the most suitable helper for effective assistance. Furthermore, in Figure \ref{case2}, we show the visualization of the seeker and helper representations, which can illustrate that the proposed method can guarantee the seniority order between the recommended helpers and the problem seekers from an end-to-end perspective.

 \section{Conclusion}
To address the urgent need for valuable healthcare information in Online Health Communities (OHCs), this study presents a seeker-helper recommender system that facilitates the recommendation of experienced helpers to assist seekers. Unlike conventional user-item recommendation approaches, our proposed framework is designed to effectively capture the social support dynamics among patients and leverage the heterogeneous information available in OHCs.
The framework incorporates a Graph-based Disentangled VAE to capture both time-invariant and time-varying information pertaining to patients. The time-invariant features represent patients' inherent knowledge, which plays a crucial role in computing similarity measures. Additionally, we introduce a monotonic regularizer to modulate the behavior of the time-varying features, ensuring their consistent progression over time. Furthermore, we enforce the logical seniority orders by incorporating a seniority-level constraint into the framework.
By considering these novel components, our proposed seeker-helper recommender system effectively addresses the unique challenges posed by OHCs, enabling the identification and recommendation of experienced helpers to respond to seekers' questions.
\bibliographystyle{IEEEtran}
\bibliography{sample-base}{}
\end{document}